\documentclass[12pt,preprint]{aastex}
 
\usepackage{graphicx}
\usepackage{epstopdf}
\usepackage{psfig}

\slugcomment{For the Astrophysical Journal}
 
\shorttitle{The anatomy of an extreme starburst group within 1.3Gyr of
the Big Bang.}

\shortauthors{Carilli et al.}
 
\begin{document}

\title{The anatomy of an extreme starburst within 1.3Gyr
of the Big Bang revealed by ALMA}

\author{C.L. Carilli\altaffilmark{1,2}, D. Riechers\altaffilmark{3},
F. Walter\altaffilmark{4},R. Maiolino\altaffilmark{2},
J. Wagg\altaffilmark{5}, L. Lentati\altaffilmark{2},
R. McMahon\altaffilmark{6}, A. Wolfe\altaffilmark{7}}

\altaffiltext{1}{National Radio Astronomy Observatory, P. O. Box 0,
Socorro, NM 87801, USA; ccarilli@aoc.nrao.edu} 
\altaffiltext{2}{Astrophysics Group, Cavendish Laboratory, JJ Thomson Avenue,
Cambridge CB3 0HE, UK}
\altaffiltext{3}{Astronomy Department, California Institute of Technology, 
MC 249-17, 1200 East California Boulevard, Pasadena, CA 91125, USA}
\altaffiltext{4}{Max-Planck-Institut f\"ur Astronomie, K\"onigstuhl 17, 
D-69117 Heidelberg, Germany}
\altaffiltext{5}{European Southern Observatory, Alonso de Cordova 3107, 
Vitacura, Casilla 19001, Santiago 19, Chile}
\altaffiltext{6}{Institute of Astronomy, University of Cambridge
Madingley Road Cambridge CB3 0HA, UK}
\altaffiltext{7}{Department of Physics and Center for Astrophysics and 
Space Sciences, University of California, San Diego, La Jolla, CA 92093, USA}

\altaffiltext{$\star$}{ALMA is a partnership of ESO (representing its
member states), NSF (USA) and NINS (Japan), together with NRC (Canada)
and NSC and ASIAA (Taiwan), in cooperation with the Republic of Chile. 
The National Radio Astronomy Observatory is a facility of 
the National Science Foundation operated under cooperative agreement by 
Associated Universities, Inc.
}

\begin{abstract}

We present further analysis of the [CII] 158$\mu$m fine structure line
and thermal dust continuum emission from the archetype extreme
starburst/AGN group of galaxies in the early Universe, BRI 1202-0725
at $z=4.7$, using the Atacama Large Millimeter Array.  The group is
long noted for having a closely separated (26kpc in projection)
FIR-hyperluminous quasar host galaxy and an optically obscured submm
galaxy (SMG).  A short ALMA test observation reveals a rich laboratory
for the study of the myriad processes involved in clustered massive
galaxy formation in the early Universe. Strong [CII] emission from the
SMG and the quasar have been reported earlier by Wagg et al. (2012)
based on these observations.  In this letter, we examine in more
detail the imaging results from the ALMA observations, including
velocity channel images, position-velocity plots, and line moment
images. We present detections of [CII] emission from two
Ly$\alpha$-selected galaxies in the group, demonstrating the relative
ease with which ALMA can detect the [CII] emission from lower star
formation rate galaxies at high redshift.  Imaging of the [CII]
emission shows a clear velocity gradient across the SMG, possibly
indicating rotation or a more complex dynamical system on a scale
$\sim 10$kpc.  There is evidence in the quasar spectrum and images for
a possible outflow toward the southwest, as well as more extended
emission (a 'bridge'), between the quasar and the SMG, although the
latter could simply be emission from Ly$\alpha$-1 blending with that
of the quasar at the limited spatial resolution of the current
observations.  These results provide an unprecedented view of a major
merger of gas rich galaxies driving extreme starbursts and AGN
accretion during the formation of massive galaxies and supermassive
black holes within 1.3 Gyr of the Big Bang.

\end{abstract}

\keywords{galaxies: formation, radio/FIR lines; submm: starbursts;
  physics: fundamental constants}

\section{Introduction}

The last decade has seen the solidification of two important
conclusions concerning massive galaxy formation. First is that massive
galaxies form most of their stars at early times, and the more
massive, the earlier. This point has been accentuated by the
observation of old spheroidal galaxies at $z \sim 2$ to 3, thereby
requiring active star formation at even higher redshifts (Kurk 2009;
Andreon \& Huertas-Company 2011).  Second, the observed correlation
between central black hole and spheroidal galaxy mass suggests a
causal connection between the formation of supermassive black holes
(SMBH) and their host galaxies (Haaring \& Rix 2004).

Large scale cosmological simulations show that massive galaxies and
SMBH can form at high-z via gas rich mergers, driving extreme
starbursts, and rapid accretion onto the black holes and subsequent
black hole mergers (Li et al.  2007). These process occur in the
densest, ie. most biased, regions in the early Universe.  The systems
evolve into large galaxies in rich clusters at low-z. More recent
simulations suggest that cold accretion from the IGM may also play a
role in, and possibly even dominate, the gas resupply (Khandai et al
2012). As the SMBH builds, feedback from the AGN expels gas from the
galaxy, and hinders further accretion, thereby terminating
starformation in the galaxy.

Galaxies detected in submm surveys with mJy sensitivity are important
in this regard, representing massive starburst galaxies in the distant
Universe. These include the submm galaxies (SMGs) and about 1/3 of
quasar host galaxies at $z > 2$ (Wang et al. 2011, 2010a,b; Blain et
al. 2002). These systems are hyper-luminous infrared galaxies
($L_{FIR} \sim 10^{13}$ L$_\odot$). The implied star formation rates
are $> 10^3$ M$_\odot$ year$^{-1}$ based on the FIR emission. At this
rate, a substantial fraction of the stars in the galaxy can form on
timescales $\le 10^8$ years.

The compact group of galaxies BRI 1202-0725 was one of the first $z >
4$ submm-bright systems discovered (Isaak et al. 1994), and remains
the archetype for major starbursts in gas rich mergers in the early
Universe.  The system includes an optically selected, broad absorption
line quasar with M$_{BH} \sim 10^9$ M$_\odot$ at a redshift of $z =
4.695$, and an optically-undetected SMG located 4$"$ (26 kpc)
northwest of the quasar at $z= 4.692$ (Omont et al. 1996 a,b; Hu,
McMahon, Egami 1996).  Both sources are hyper-luminous FIR galaxies
(Omont et al. 1996a,b; Iono et al. 2006; Yun et al. 2000). Both
sources have been detected in CO line emission, with implied gas
masses $\rm M(H_2) \sim$ 5$\times 10^{10}(\alpha/0.8)$ M$_\odot$
(Salome et al. 2012; Carilli et al, 2002; Ohta et al. 1996; Omont et
al. 1996b), assuming a CO luminosity to gas mass conversion factor,
$\alpha = 0.8$ M$_\odot$ (K km s$^{-1}$ pc$^{2}$)$^{-1}$, appropriate
for star forming galaxies (Solomon \& Downes 1998). Both source are
detected in radio continuum emission, consistent with either a weak
AGN or extreme star formation (Yun \& Carilli 2002). Very high
resolution CO and radio continuum imaging shows resolved emitting
regions on scales $\sim 0.3''$, likely corresponding to compact
starbursts in both galaxies on scales $\sim$ 2 kpc (Carilli et
al. 2002; Momjian et al 2005; Yun \& Carilli 2002).

The best CO observations of the BRI 1202-0725 system to date are by
Salome et al. (2012).  They find that the SMG appears extended, with
two components separated by $\sim 0.5"$ east-west. The CO emission
from the quasar also appears extended to the south by about $2"$. The
CO excitation in both galaxies is high, suggesting warm (40K to 50K),
dense ($> 10^3$ cm$^{-3}$) gas.

Extended Ly$\alpha$ emission has been seen around the quasar, with a
tail of emission extending from the quasar toward the SMG, terminating
in a Ly$\alpha$ emitting galaxy about 2.3$"$ northwest of the quasar
(Hu et al. 1996; Ohta et al. 2000; Ohyama et al. 2004). This galaxy is
designated Ly$\alpha$-1 by Salome et al. (2012).  Spectra of this
galaxy yield a redshift of $4.702\pm 0.009$ (Petitjean et al.1996).
The large uncertainty in the redshift reflects the fact that the
Ly$\alpha$ profile for this galaxy is broad and complex.  Continuum
emission from Ly$\alpha$-1 is detected in the HST i-band image (Hu et
al. 1996).  Klamer et al.  (2004) discuss the possibility of
jet-induced star formation in this system, although an extended radio
jet has not been detected to date.  A second galaxy has been detected
in Ly$\alpha$ emission, and in the HST i-band image, located $2.7''$
southwest of the quasar (Hu et al. 1996; 1997; Fontana et al. 1996).
Salome et al. (2012) designate this galaxy Ly$\alpha$-2.  Both of
these galaxies appear extended on a scale $\sim 0.5"$ in the HST
images, toward the quasar.

In an earlier letter, Wagg et al. (2012) presented an initial analysis
of the [CII] and dust continuum emission from the SMG and the quasar
host galaxy in BRI1202-0725 using a test observation from ALMA.  The
[CII] 158$\mu$m line is typically the strongest emission line from
cool gas in star forming galaxies, and is a principle ISM gas coolant,
tracing both photon-dominated regions enveloping active star forming
clouds, low density ionized gas, and the cold neutral medium (Genzel
\& Cesarsky 2000; Stacey et al. 2010; Malhotra et al.  1997; Crawford
et al. 1986).  As such, the [CII] line is a key diagnostic of ISM
energetics in early galaxies as well as of galaxy dynamics (Carilli \&
Walter 2013).  Wagg et al. (2012) find that the [CII]/FIR luminosity
ratio is relatively low in both the SMG and quasar, $8\times 10^{-4}$
and $2 \times 10^{-4}$, respectively. For comparison, the Milky Way
has a value of $3\times 10^{-3}$ (Wright et al. 1991; Bennett et
al. 1994). Low [CII]/FIR ratios are typically seen in extreme
starburst environments, and may arise due to the high radiation
environment leading to substantially charged dust grains, and hence
less efficient gas heating via photoelectric emission from the dust
grains (Malhotra et al. 1997). Alternatively, Sargsyan et al. (2012)
propose that a substantial contribution to dust heating by an AGN may
lead to low ratios in very luminous systems, while Papadopoulos et
al. (2010) suggest that optical depth effects at 158$\mu$m may also
play a role.

In this paper, we present a more detailed analysis of the ALMA data,
with a closer look at the imaging results based on self-calibration of
the visibility data.  We present spectra, channel images, moment
images, and a position velocity analysis. The [CII] imaging reveals a
rich physical environment. Both Ly$\alpha$ emitting galaxies are also
detected in [CII] emission. The SMG shows a clear velocity gradient in
the atomic gas on a scale $\sim 10$kpc, and the quasar shows evidence
for atomic gas outflow, or a possible tidal feature to the southwest
toward Ly$\alpha$-2.  There may even be a bridge of cool gas connecting
the quasar and the SMG, although the presence of Ly$\alpha$-1 confuses
this possibility. 

\section{Observations}

Observations were made of BRI 1202-0725 with the Atacama Large
Millimeter Array during science testing and verification in January
2012 (see Wagg et al. 2012 for more details).  These short
observations (25 minutes on-source), with only 17 antennas, are
already an order of magnitude more sensitive than any previous submm
line or continuum observations of the system.  Observations were made
with 4 dual-polarization bands of 2GHz bandwidth each.  The first band
covered the [CII] line from the SMG and the quasar, with a central
(sky) frequency of 333.960 GHz, corresponding to a redshift of $z =
4.6909$ for the [CII] 158$\mu$m line (rest frequency = 1900.539
GHz). The other bands were centered off the line by $\sim 10$ GHz, and
these were used for continuum imaging. Hanning smoothing was applied
on output from the correlator to the 128 channel per 2 GHz band
spectral data.

Calibration of the data was performed using the Astronomical Image
Processing System (AIPS). The source
3C279 was used as a flux density and phase calibrator, using the flux
density at 340GHz of 17Jy derived by Wagg et al. (2012). 3C279 is
located 13$^o$ from the target source.  A 10min calibration cycle was
employed during the observations, and the weather conditions were
excellent, with gains stable to a few degrees and a few percent over
the hour. The source 3C279 was also used for bandpass calibration, and
the bandpass was smooth and stable, again to within a few percent and
a few degrees except for a few edge channels in each band that were
flagged. 

The sensitivity of even the reduced array of ALMA is such that
self-calibration could be employed using the continuum and line
emission from BRI 1202-0725 (total continuum flux density
of 37mJy at 340GHz). Amplitude and phase self-calibration was
performed using a 3min averaging time for complex gain solutions.

Imaging was performed using a robust imaging scheme with R=1 (Cornwell
et al. 1999), for which the restoring Gaussian beam was
$1.2''\times 0.8''$ resolution (major axis North-South).  For the
spectral line data, the continuum was subtracted from the UV data
using a CLEAN component model derived from the three off-line bands,
before imaging and deconvolution.

\section{Results}

Figure 1a shows the 340 GHz continuum emission from BRI 1202-0725,
overlayed on the HST i-band image. For relative astrometry
(optical-radio) we follow Carilli et al. (2002) by aligning the quasar
position in the HST, Ly$\alpha$, submm continuum, and radio
continuum images.  Given the scatter in these positions, a rough
estimate of the relative astrometry is $\sim 0.25''$. 

The continuum emission from the SMG and quasar is unresolved
(deconvolved sizes $< 0.3"$ derived from Gaussian fitting), with flux
densities of $17\pm 2.5$mJy and $18\pm 2.7$mJy, respectively,
including the 15\% uncertainty in absolute gain calibration estimated
by Wagg et al. (2012). The galaxy Ly$\alpha$-2 to the southwest of the
quasar is also detected in the continuum, and is unresolved with a
flux density $1.4\pm 0.2$ mJy (see also Wagg et al. 2012). This submm
continuum source peaks $0.4''$ from Ly$\alpha$-2 as seen on the HST
image (Hu et al. 1996), although this galaxy appears extended in the
HST image in the direction of the submm source and the quasar by at
least $0.5''$ (Figure 1a).
 
Figure 2 shows the channel images of the [CII] line, smoothed to a
spectral resolution of 62.5 MHz (= 56 km s$^{-1}$).  Spectra of the
sources are shown in Figure 3 at 31.25 MHz spectral resolution (= 28
km s$^{-1}$). 

Strong [CII] emission is seen from the SMG over essentially the full
frequency range in the first band.  Likewise, strong emission is seen
from the quasar over a narrower frequency range.  For reference, we
have fit Gaussian models to the spectra, although admittedly the SMG
spectrum is decidedly non-Gaussian. Results from the fitting are given
in Table 1 (see Wagg et al. 2012 for more details on the SMG and
quasar spectra).

Inspection of the [CII] channel images shows emission from the
vicinity of Ly$\alpha$-1, between the quasar and the SMG, in a few
channels just above 0 km s$^{-1}$.  Figure 1b shows the integrated
emission in the velocity range 0 km s$^{-1}$ to 100 km s$^{-1}$, with
the Ly$\alpha$ image of Hu et al. (1996) in color. This apparent [CII]
emission is seen along the direction of the extended Ly$\alpha$ tail
from the quasar toward the SMG (Hu et al. 1997), peaking close (within
$0.3"$) to galaxy Ly$\alpha$-1. A spectrum at this peak position near
Ly$\alpha$-1 is shown in Figure 3c.  The [CII] emission from
Ly$\alpha$-1 is narrow, with a FWHM $\sim 56$ km s$^{-1}$, centered at
$z = 4.6950$. Note that it is difficult to conclude based on these
relatively low spatial resolution data whether this emission is truly
extended between the quasar and Ly$\alpha$-1, or simply blended by the
limited spatial resolution.  The [CII] emission around the quasar also
appears extended in the direction of Ly$\alpha$-2 in the channels
around +250 km s$^{-1}$ to +300 km s$^{-1}$.

The two highest velocity channels in Figure 2 show [CII] emission from
galaxy Ly$\alpha$-2. Figure 3d shows the spectrum of this galaxy.
Since the redshift of this galaxy was essentially unknown prior to the
observations, the spectral coverage truncates some unknown fraction of
the line at high velocity (low frequency).  Hence, we do not know the
full velocity extent nor total luminosity of the [CII] emission for
Ly$\alpha$-2, and the redshift is strictly a lower limit.

Figure 4 shows the iso-velocity image (moment 1 = intensity weighted
mean velocity) of the [CII] emission made using a 3.5$\sigma$
blanking level.  There is a clear east-west velocity gradient across the SMG. 
This can be seen in the channel images (Fig 2), where the peak of 
the line emission moves by
about $1.7"$ east to west from low to high velocities. The tail of 
emission to the north of the quasar toward Ly$\alpha$-1 is also
seen, as well as the extended emission to the
southwest of the quasar toward Ly$\alpha2$. 

Figure 5 shows a position-velocity (PV) plot along the major axis
(East-West) of the SMG. The E-W velocity gradient is clear, as well as
a twist in the line-of-nodes at the extrema in velocity. Such a twist
could signal a warped-disk in atomic gas.
 
\section{Analysis}

Wagg et al. (2012) present a detailed analysis of the main [CII] line
and dust continuum emission from the quasar and SMG in BRI 1202-0725. 
Herein we focus on the [CII] imaging, and in particular, the extended emission
from the SMG, quasar, and the Ly$\alpha$ emitting galaxies. 

\subsection{SMG dynamics}

The isovelocity contours of the [CII] emission from the SMG show a gradient
of 525 km s$^{-1}$ over $\sim 1.7''$ (Figure 4).  If due to simple
rotation, this would imply a dynamical mass of $3.6\times 10^{11}$
[sin(i)/sin(45)]$^{-2}$ M$_\odot$ to a radius of 5.5 kpc.  This mass
is an order of magnitude higher than the molecular gas mass (Carilli
et al. 2002), unless one adopts the five times larger CO luminosity to
gas mass conversion factor for the Milky Way relative to the nuclear
starburst value. However, recent observations suggest that a starburst
conversion factor is clearly favored in SMGs (see Carilli  \& Walter 2013
for a review).

Perhaps the simplest conclusion is that the velocity gradient in the
SMG is not due to rotation. Salome et al. (2012) find that the CO 5-4
line profile for the SMG appears more like a double-line system as
opposed to the more gradual gradient seen to higher velocities in the
[CII] line spectrum. Likewise, the east-west PV for the SMG in CO
5-4 shows a more broken distribution, with a dip, then plateau to high
velocity.  The [CII] shows a more continuous distribution.  Salome et
al.  conclude that the SMG itself may in fact be an end-stage merger
of two gas rich galaxies.

\subsection{Quasar outflow}

The quasar [CII] emission line is well fit by a Gaussian, except at
the high velocity end of the spectrum, where there appears to be a
broad wing. Figure 3 shows a double Gaussian fit to the spectrum
including this broad wing. In this case, the main component has a
redshift of $4.6941$, a peak of 24mJy, and a FWHM = 274 km s$^{-1}$. 
The second component for the wing has a peak flux
density of 2.4 mJy, centered at $z=4.6967$, or 141 km s$^{-1}$ from
the quasar redshift, and a FWHM = 395 km s$^{-1}$ (Fig 2b). Inspection
of the channel images (Fig 2) shows emission extending south of the
quasar in this velocity range.  The deconvolved size of the emission
in this velocity range is $\sim 1.5''$ oriented north-south, as
derived from Gaussian fitting. This feature could be the result of the
tidal interaction, or an outflow. We briefly consider the latter
possibility.

In order to determine the outflow rate and energetics, we follow the
same approach presented in Maiolino et al. (2012), where they detect a
similar outflow traced by [CII] in a z=6.4 quasar.  More specifically,
the luminosity of the high velocity gas traced by the [CII] provides a
lower limit on the mass of outflowing atomic gas (in the extreme limit
$n \gg n_c$, the critical density for [CII] excitation). Since the
broad component of [CII] is poorly constrained, we conservatively
consider only the luminosity of the [CII] wing integrated between +220
km/s and +500 km/s ($\rm L_{[CII]}(wing)=0.3\times 10^9 L_{\odot}$),
which gives a lower limit on the outflowing atomic gas mass of $\rm
M_{outfl}(atomic) > 2.9\times 10^8 M_{\odot}$, assuming an abundance
ratio of $\rm [CII]/H = 1.4\times 10^{-4}$ (Savage \& Sembach 1996).
Assuming a simplified conical or bi-conical outflow geometry and a
size of 1.5$''$ (=10 kpc), we infer a lower limit on the outflow rate
of $\rm \dot{M}_{outfl} > 80 M_{\odot}~ yr^{-1}$. This is a modest
outflow rate relative to other quasar-driven outflows, which generally
have outflow rates higher than a few hundred up to a few thousands
$\rm M_{\odot}~ yr^{-1}$ (Feruglio et al. 2010, Sturm et al. 2011,
Cano-Diaz et al. 2012, Maiolino et al. 2012). However, one should keep
in mind that this is a lower limit on the total outflow rate, because
of the assumed extreme physical conditions, and because a significant
fraction of the outflow may be in the molecular phase.  We note that
by taking the flux of the full broad component used to fit the broad
wing (Fig.3a) the inferred outflow rate would increase by a factor of
about three. Moreover, this calculation depends critically on the
assumed conical geometry and constant velocity, and hence should be
considered roughly order of magnitude (see Maiolino et al. 2012 for
details).

We derive a lower limit on the kinetic power of $\rm P_K(outfl) >
5.3x10^{42} erg~ s^{-1}$, which is only 0.2\% of the quasar bolometric
luminosity ($\rm 2.6\times 10^{45} erg~ s^{-1}$), well below the
maximum kinetic power that can be driven by the quasar according to
models ($\sim$5\%, Lapi et al. 2006), meaning that the quasar
radiation pressure can easily drive such a mild outflow.  On the other
hand, this relatively low kinetic power can also be driven by the
starburst hosted in the quasar itself. The total IR luminosity of the
quasar is $3.3\times 10^{13}$ L$_\odot$ (Wagg et al. 2012), which
implies a star formation rate of $\rm \sim 3300 M_\odot~ yr^{-1}$, for a
Charbrier IMF using standard relations (Carilli \& Walter
2013). The expected kinetic power is $\rm P_K(SB)\sim 2\times 45~ erg~
s^{-1}$ (Veilleux et al. 2005).  Therefore, in this case the
energetics cannot constrain the nature of the outflow: both the quasar
and the starburst can drive it. And again, the derived kinetic
luminosity is strictly a lower limit.

\subsection{The Ly$\alpha$ emitters}

We have detected [CII] line emission from two narrow band selected,
Ly$\alpha$ emitting galaxies in the BRI 1202-0725 group at $z
=4.7$. The optical and near-IR properties of Ly$\alpha$-1 and 2 are
described in Hu et al. (1996). The I-band magnitudes are 24.2 and
24.5, respectively. Hu et al. (1996) conclude that these galaxies have
magnitudes and colors 'roughly comparable to an $L_*$ galaxy [at high
redshift]' (see also Hu et al. 1997; Lu et al. 1996; Fontana et
al. 1996). They also conclude that the Ly$\alpha$ equivalent widths
are: ' consistent with excitation by the underlying [stellar]
population', although they note that the proximity of the quasar
raises some interesting questions about ionization (see below).  We
first consider the submm line and continuum properties of these in the
context of star formation.

For Ly$\alpha$-2, the FIR luminosity is $1.7\times 10^{12}$ L$_\odot$,
based on the 340GHz flux density, and the implied star formation rate
derived from the FIR luminosity is 170 M$_\odot$ year$^{-1}$, assuming
a standard dust SED and conversion factors (Carilli \& Walter 2013).
The [CII] line luminosity is $8.2\times 10^8$ L$_\odot$, based on the
symmetrized Gaussian fit shown in Figure 2 (again, this is highly uncertain
due to the truncated line profile).  Sargsyan et al. (2012)
present an empirical relationship between [CII] luminosity and star
formation: log(SFR) = log(L(CII)) - 7.08 (solar units), derived from
observations of the PAH, FIR, and [CII] emission from a large sample
of nearby galaxies.  For Ly$\alpha$-2 this relationship would imply a
star formation rate of 70 M$_\odot$ year$^{-1}$.

The [CII]/FIR ratio is 0.0005 for Ly$\alpha$-2. This ratio is
comparable to nuclear starburst galaxies seen nearby, and well below
the Milky Way ratio of 0.003.  However, we emphasize that the [CII]
luminosity is uncertain due to the truncated line profile seen in
Figure 3, and hence these estimates should be considered at best
order-of-magnitude.

For Ly$\alpha$-1, the [CII] line luminosity is $1.9\times 10^8$
L$_\odot$, based on the Gaussian fit shown in Figure 2, with an
implied star formation rate of 19 M$_\odot$ year$^{-1}$, using the
Sargsyan et al. (2012) relationship.  In this case, we only have lower
limit to the FIR luminosity of $3.6\times 10^{11}$ L$_\odot$, based on
an upper limit to the 340 GHz flux density of 0.3mJy, or an upper
limit to the star formation rate of 36 M$_\odot$ year$^{-1}$.  In this
source, the lower limit to the [CII]/FIR ratio is 0.0005.

The proximity of a very luminous quasar to these Ly$\alpha$ emitting
galaxies ($L_{1450} \sim 10^{13}$ L$_\odot$), presents the possibility
that atomic gas is both heated and ionized by the luminous nearby
quasar.  We ran a series of simple CLOUDY models (Ferland et al 1998),
that reproduced both its Lyman-alpha and [C II] luminosities in
Ly$\alpha$-1.  We adopted simple models in which a small segment of a
spherical shell of gas is located a radial of distance 15 kpc from the
quasar. We explored an initial density range between 100 and 1000
cm$^{-3}$, but then pressure equilibrium was assumed between adjacent
cells with an initial temperature of 10000 K. Thus density climbed to
higher values in colder, neutral regions beyond the ionization
front. Densities much lower than this required unreasonably high
spatial extent of the gas.  Densities much higher resulted in very
small cloud radial dimensions compared to larger lateral dimensions
required by the [CII] 158$\mu$m emission maps.

The gas exhibits the classic ``H II'' region zone facing the quasar
that extends into an ionization front further away, followed by a zone
of neutral gas even further away from the quasar. In every case the
Ly-alpha emission arose in the warm $T{\sim}10^{4}$ K, ionized gas,
while the 158$\mu$m emission arose in the neutral gas.  But at the
same time other standard UV emission lines such as Si IV 1397, N V
1239, and OI 1403 were predicted at luminosities more than a factor of
10 higher than the upper limits placed by Ohyama etal. (2004). While
these authors did not observe C IV$\lambda$1548, which is the
strongest of these transitions, neither has it been detected by other
less sensitive observations, which are clearly in conflict with its
predicted luminosity of $\sim 10^{42}$ ergs s$^{-1}$ (Hu et
al. 1997). Lowering the metal abundance does not help since many of
these transitions serve as important coolants and the gas adjusts to
higher temperatures to balance the heat input.

We conclude that the quasar is an unlikely source of heat and
ionization for Galaxy Ly$\alpha$-1. As a result, we speculate that a
torus that is optically thick to ionizing photons shields the quasar
in the direction of Ly$\alpha$ galaxies (Goosmann \& Gaskell 2007). We
note that, if the FUV heating the PDR arises from the quasar, then the
high value of the FUV intensity is likely to produce [OI] 63$\mu$m
emission at fluxes comparable or exceeding the [CII] 158$\mu$m
emission. The point here is that at 15 kpc distance, the quasar will
produce a $G_0 \sim 1000$, where $G_0$ is the FUV energy density in
units of the Habing ISM value ($G_0 = J_\nu/(10^{-19}$ ergs cm$^{-2}$
s$^{-1}$ sterad$^{-1}$ Hz$^{-1}$)).  As Kaufman et al. (1999) have
shown, PDRs with $G_0 > 1000$, result in an intensity ratio 
[OI] 63$\mu$m/[CII] 158$\mu$m $> 1$, owing to the relatively high
temperature of the gas. Thus the detection of the 63$\mu$m transition
at fluxes comparable or greater than that of the 158$\mu$m line
indicates a relatively high value of $G_0$.  This would be consistent
with incidence FUV flux coming from the quasar, or local
heating by a luminous O star located within a few pc from the
gas. An intensity ratio $<< 1$, would certainly rule out the gas
irradiated by such an intense radiation field.

\section{Discussion}

BRI 1202-0725 shows all the attributes expected during massive galaxy
and SMBH formation in a dense group in the early Universe. The system
has four interacting galaxies within 5'' of each other (35 kpc in
projection): two hyper-starbursts (an SMG and a quasar host), and two
Ly$\alpha$ emitting galaxies (Ly$\alpha$-1 and 2).  Impressively, all
are detected in [CII] emission. 

Perhaps the most impressive result from these observations is the
relative ease with which high redshift Ly$\alpha$-selected galaxies
can be be detected in [CII] emission. Using only 1/3 of the full ALMA,
both Ly$\alpha$-1 and 2 were detected at reasonable signal to noise in
just 20min.  Scaling to the full ALMA, and a still reasonable 3hr
integration, leads a factor 10 increase in sensitivity, allowing for
detailed imaging of the atomic gas on sub-kpc scales in normal star
forming galaxies (star formation rates $\sim$ few to 10 M$_\odot$
year$^{-1}$), at very high redshift. [CII] observations with ALMA open
a new window on dynamical studies of the first galaxies.

An interesting trend for the [CII] emission in the vicinity of the
quasar is for the atomic gas to extend in the direction of
Ly$\alpha$-1 and 2. In fact, Ly$\alpha$-1 may be a local maximum within a
possible bridge of atomic gas connecting the quasar and the SMG, as
expected during major mergers of gas rich galaxies (Barnes \&
Hernquist 1996; Li et al. 2008). Unfortunately, the current resolution
is insufficient to distinguish between a true bridge, or simply
blurring of independent [CII] emission from the quasar and G3. An
argument for some physical connection is the velocity continuity from
the quasar through G3. A critical observation will be to confirm
this bridge with deeper, higher resolution observations with ALMA, as
well as to confirm the [CII] line and detect dust continuum emission
from Ly$\alpha$-1.

The presence of a luminous quasar would naturally lead to strong
outflows, as has been seen in both low and high redshift quasars in CO
and/or [CII] emission (Maiolino et al. 2012; Sturm et
al. 2011). Outflow from the quasar in BRI 1202-0725 is suggested by
the presence of an associated broad absorption line system (Capellupo
et al. 2011), although this could also be of tidal origin.  Our
analysis of the broad [CII] wing of the quasar indicates that, while
an outflow may be present, it is not likely to dominate the gas
depletion in the quasar host galaxy. Indeed, at the current outflow
rate the available gas mass ($\rm \sim 5\times 10^{10} M_\odot$) would
be expelled in about $\rm 6\times 10^8 yr$, which is much longer than
the star formation consumption timescale ($\sim 10^7$ yr).

The [CII] emission from the SMG shows a clear, relatively smooth
velocity gradient over a scale of $\sim 1"$ in an east-west PV plot,
and in the first moment image, and a smooth integrated emission line
profile over a broad range in velocity (700 km s$^{-1}$). The
morphology seen in the east-west PV plot, and the spectral profile,
for the [CII] emission differ from the CO 5-4 emission, which appears
more 'clumpy' (two distinct peaks) in both the PV plot, and in the
spectrum (Sargsyan et al. 2012).  High order CO emission traces higher
density gas, $> 10^3$ cm$^{-3}$, while [CII] traces both the high
density gas in PDRs, and lower density atomic gas (CNM).  Hence, it
could be that the lower density atomic gas has a smoother overall
distribution than the CO, filling-in the velocity and spatial gaps
seen in CO. Higher resolution CO and [CII] observations are required
to delineate the relative atomic and molecular gas distributions in
BRI 1202-0725.

Simpson et al. (2012) recently presented evidence that FIR luminous
quasars are a short (1 Myr) but ubiquitous phase during the evolution
of a dust obscured gas rich starburst to an unobscured, gas poor
quasar.'  The narrower CO line width in the quasar relative to the SMG
in BRI1202-0725, and the apparently more spatially extended gas
distribution (as seen from the dynamics) in the SMG, are consistent
with the trends discussed by Simpson et al.. However, in BRI1202-0725
we see the quasar and SMG phases as coeval, with both galaxies
involved in the gas-rich merger. It is unclear how such a
hyper-starburst galaxy pair, one of which is a luminous broad line
quasar, fit-in to the stated evolutionary model.

\acknowledgements CLC thanks the Kavli Institute for Cosmology for
their hospitality. We thank the referee for useful comments. This
paper makes use of the following ALMA data:
ADS/JAO.ALMA\#2011.0.00006.SV. ALMA is a partnership of ESO
(representing its member states), NSF (USA) and NINS (Japan), together
with NRC (Canada) and NSC and ASIAA (Taiwan), in cooperation with the
Republic of Chile. The Joint ALMA Observatory is operated by ESO,
AUI/NRAO and NAOJ. The National Radio Astronomy Observatory is a
facility of the National Science Foundation operated under cooperative
agreement by Associated Universities, Inc.

\clearpage
\newpage

Andreon, S. \& Huertas-Commpany, M. 2011, A\& A, 526, 11

Barnes, J, \& Hernquist, L. 1996, ApJ, 471, 115

Blain, A., Smail, I. Ivison, R., Kneib, J.P., Frayer, D.
2002, PhR, 369, 111

Bennett, C.L., Fixsen, D., Hinshaw, G. et al. 
1994, ApJ, 434, 587

Cano-Diaz, M, Maiolino, R., Marconi, A. et al. 2012, A\& A, 537, L8

Cantalupo, S., Lilly, S., Haehnelt, M. 2012, MNRAS, in press

Capellupo, D. M., Hamann, F., Shields, J. C. et al. 2011,
MNRAS, 413, 908

Carilli, C.L, Kohno, K, Kawabe, R. et al. 2002, AJ, 123, 1838

Carilli, C.L, Walter, F., Riechers, D. et al. 2011,
Rev. Mod. Astro., 23, 131

Carilli, C.L. \& Walter, F. 2013, ARAA, in press 

Carilli, C.L., Hodge, J., Walter, F. et al. 2011, ApJL,
739, 33

Christensen, Sanchez, S., Jahnke, K., et al. 2004, A\& A,
417, 487

Cornwell, T., Braun, R., Briggs, D. 1999, ASP Conference
Series, Vol. 180, Synthesis Imaging in Radio Astronomy II, eds
G. B. Taylor, C. L. Carilli, and R. A. Perley, p. 151

Crawford, M.K., Lugten, J., Fitelson, W., Genzel, R., Melnick, G. 1986, ApJ,
303, L57

Fontana, A. et al. 1996, MNRAS, 279, L27

Ferland, G. J. Korista, K.T. Verner, D.A. Ferguson,
J.W. Kingdon, J.B. Verner, \& E.M. 1998, PASP, 110, 761

Feruglio, C., Maiolino, R., Piconcelli, E. et al. 2012, A\& A, 518, L155

Genzel, R. \& Cesarsky, C. 2000, ARAA, 38, 761

Goosmann, R, Gaskell, C. 2007, A\& A, 465, 129

Haaring, N. \& Rix, H. 2004, ApJ,  604, L89,

Hibbard, J., van der Hulst, J., Barnes, J., Rich, M. 2001, AJ, 122, 2696

Hodge, J., Carilli, C., Walter, F. et al. 2012, ApJ, 
in press

Hu. E. M, McMahon, R.G., Egami, E. 1996, ApJ, 459, L53

Hu. E. M, McMahon, R.G., Egami, E. 1997, Proceedings of
the 37th Herstmonceux conference, p. 91, eds.  Nial R. Tanvir, Alfonso
Aragon-Salamanca, and Jasper V. Wall (Singapore: World Scientific)

Isaak, K., McMahon, R.G., Hills, R.E., Withington,
S. 1996, MNRAS, 269, L28

Israel, F. P., Maloney, P. R. 2011, A\& A, 531, 19

Iono, D., Yun, M.S., Elvis, M., et al. 2007, 645, L97
 Levshakov, S. et al. 2008, A\& A, 479, 719

Khandai, N. et al. 2012, MNRAS, 423, 2397


Klamer, I., Ekers, R., Sadler, E. et al. 2004, ApJ 612, L97

Kurk, J. 2009, A\& A, 504, 331

Lapi, A., Shankar, F., Mao, J. et al. 2006, ApJ, 650, L42

Li, Y., Herquist, L., Robertson, B. et al. 2007, ApJ, 665, 187

Lu, L. et al. 1996, ApJ, 457, 1

Malhotra, S., Helou, G., Stacey, G., et al. 1997, ApJ, 491, L27

Maiolino, R., Gallerani, S., Neri, R. et al. 2012, MNRAS letters in press

Omont, A., Petitjean, P., Guilloteau, S., et al. 1996,
Nature, 382, 428

Ohta, K., Matsumoto, T., Maihara, T., Iwamuro, F. et al. 2000,
PASJ, 52, 55

Ohta, K., Yamada, T., Nakanishi, K., Kohno, K., Akiyama,
M., Kawabe, R. 1996, Nature, 382, 426

Ohyama, Youichi; Taniguchi, Yoshiaki; Shioya, Yasuhiro
2004, AJ, 128, 2704

Papadopoulos, P., Isaak, K., van der Werf, P. 2010,
ApJ, 711, 757

Petitjean, P., Pecontal, E., Valls-Gabaud, D., Charlot,
S. 1996, Nature, 380, 411

Sargsyan, L., Lebouteiller, V., Weedman, D. et al. 2012, ApJ, submitted

Savage, B. \& Sembach, K. ARAA, 43, 677

Solomon, P. \& Downes, D. 1998, ApJ, 507, 615

Stacey, G. et al. 2010, ApJ 725, 957

Sturm, E., González-Alfonso, E., Veilleux, S. et al. 2011, ApJ, 733, L16

Tacconi, L., Genzel, R., Smail, I. et al. 2008, ApJ, 680, 246

Veilleux, Sylvain, Cecil, Gerald, Bland-Hawthorn, Joss
2005, ARAA, 43, 769

Wagg, J. et al. 2012, ApJ, submitted

Walter, F. \& Carilli, C.L. 2007, HiA, 14, 263

Wang, R., Wagg, J., Carilli, C. et al. 2011, AJ, 142, 101

Wang, R., Wagg, J., Carilli, C. et al. 2010a, ApJ, 739, L34

Wang, R., Carilli, C., Neri, R. et al. 2010b, ApJ, 714, 699

Wright, E., Mather, J., Bennett, C. et al. 1991, ApJ, 381, 200

Yun, M.S., Carilli, C.L., Kawabe, R. et al. 2000, ApJ, 528, 171

Yun, M.S. \& Carilli, C. 2002, ApJ, 568, 88

\clearpage
\newpage

\begin{table}\label{Observed}
\begin{center}
\caption{BRI 1202-0725 parameters}
\begin{tabular}[ht]{|c|c|c|c|c|c|}
\tableline
Source & Position$^a$ & S$_{\rm 340GHz}$ & 
[CII] Redshift$^b$ & Line Peak & FWHM$^c$ \\
~ & J2000 & mJy & ~ & mJy & km s$^{-1}$ \\
\tableline
Quasar$^d$ & J120523.13-074232.6 & $18\pm 2.7$ & 4.6942 & $24 \pm 3.6$ 
& 275 \\
SMG  &   J120522.98-074229.5 & $17\pm 2.5$ &  4.6915 & $16 \pm 2.4$ & 712 \\
Ly$\alpha$-1   &   J120523.06-074231.2 &  $< 0.3$ & 4.6950 & $5.1\pm 0.7$  & 56 \\
Ly$\alpha$-2$^e$   &   J120523.04-074234.3 & $1.4\pm 0.2$ & 4.7055 & $3.7\pm 0.6$ & 338 \\
\tableline
\end{tabular}
\end{center}
\noindent $^a$Submm continuum position, except for Ly$\alpha$-1, which uses
the [CII] line peak position. \\
\noindent $^b$Formal errors from Gaussian fitting 
on the redshifts are $< 0.0003$ in all cases. \\
\noindent $^c$Formal errors from Gaussian fitting 
on the line widths are $\le 15$ km s$^{-1}$. \\
\noindent $^d$Redshift for single Gaussian fit (see section 3.1). \\
\noindent $^e$This line is truncated in velocity, and hence the fit is highly 
uncertain, and the redshift is strictly a lower limit. \\
\end{table}

\clearpage
\newpage

\begin{figure}
\psfig{file=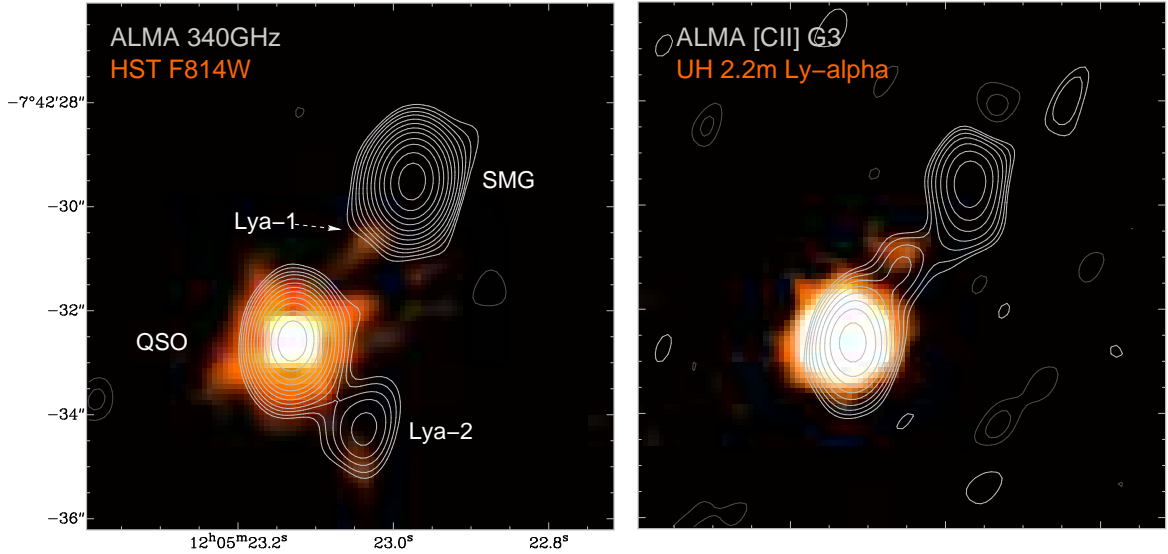,width=7in}
\caption{
{\bf Left:}
White contours show the submm continuum emission from BRI 1202-0725 at 340 GHz 
at $1.2''\times 0.8''$ resolution (major axis North-South). 
Contour levels are a geometric progression in square root
two, starting at 0.25 mJy beam$^{-1}$. Negative contours are dashed.
The rms noise on the image is 0.1mJy beam$^{-1}$. The color
shows the HST F814W image of Hu et al. (1996).
The SMG, quasar, Ly$\alpha$-1, and Ly$\alpha$-2  are indicated.
{\bf Right:} The integrated [CII] emission 
in the velocity range 0 km s$^{-1}$ to 130  km s$^{-1}$ 
(see Fig 2; Fig 3c).
Contour levels are a geometric progression in square root two, starting
at 1.0 mJy beam$^{-1}$.  The color is the Ly$\alpha$ narrow band 
image from Hu et al. (1996).  The Ly$\alpha$ image within  
$\sim 1''$  of the quasar  is saturated.
Note that saturation and diffraction spikes affect the HST and Ly$\alpha$
images within about a 1$''$ radius of the quasar.
}
\end{figure}

\begin{figure}
\psfig{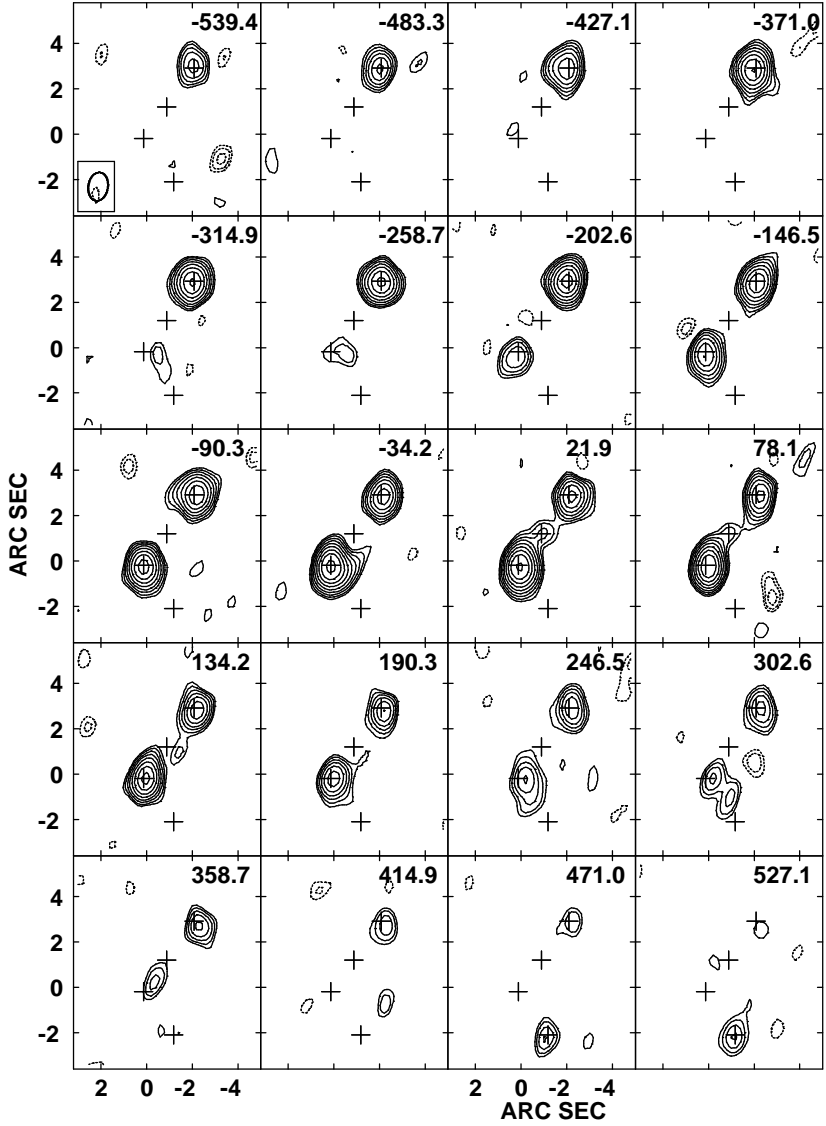}
\caption{
The [CII] channel images at 62.5MHz (= 56 km s$^{-1}$) per channel,
and $1.2''\times 0.8''$ resolution (major axis North-South).  The
channels are labeled in velocity (km s$^{-1}$), where
zero velocity
corresponds to $z = 4.6941$. 
Contour levels are a
geometric progression in square root two, starting at 1.5 mJy
beam$^{-1}$. Negative contours are dashed.  The crosses show the
positions of the peak in the submm continuum emission from the SMG, quasar, and
Ly$\alpha$-2, and the [CII] peak in Ly$\alpha$-1. The continuum derived from the off-line bands has been
subtracted.  The rms noise is 0.64mJy beam$^{-1}$ channel$^{-1}$.
}
\end{figure}

\begin{figure}
\psfig{file=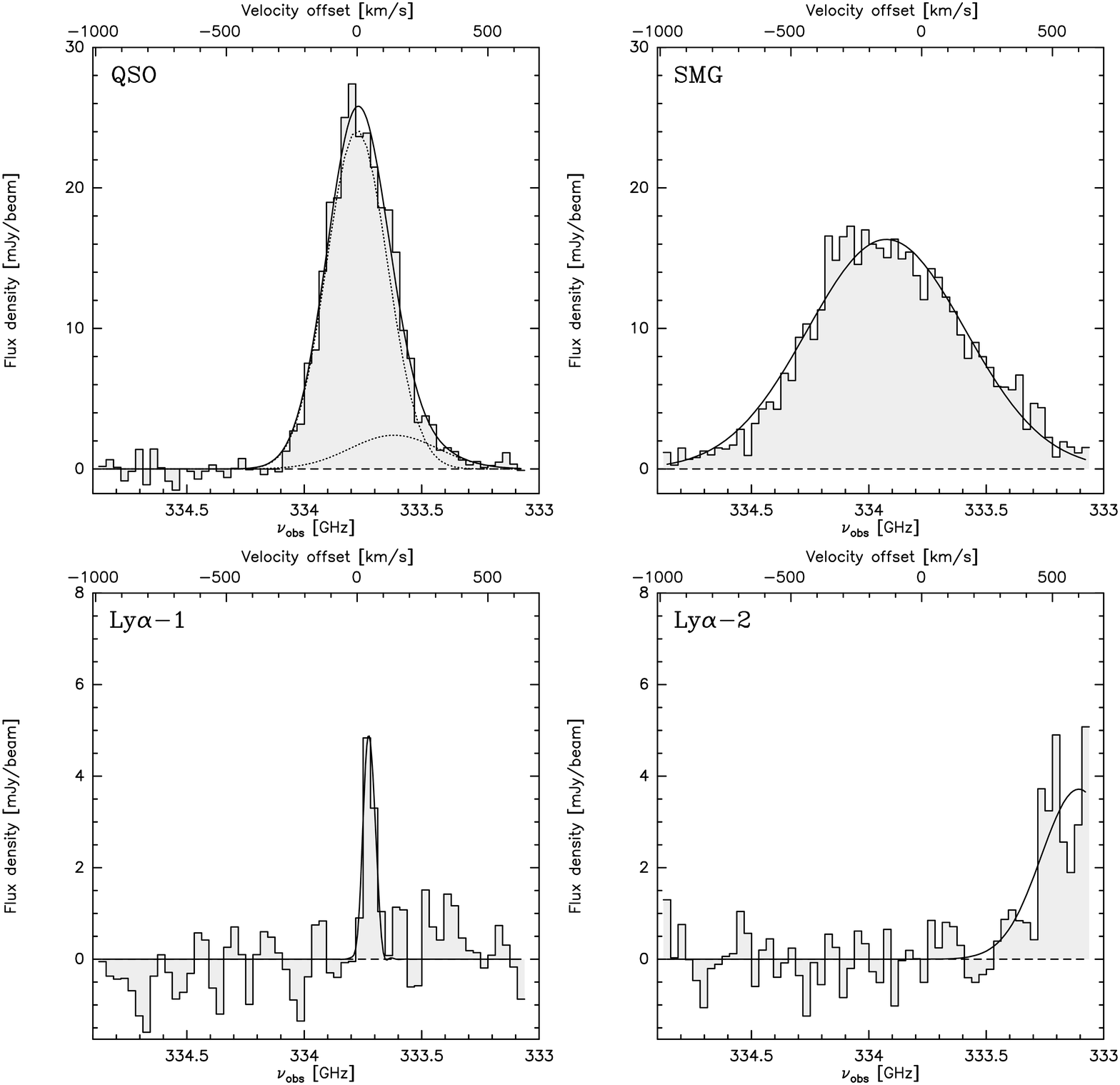,width=6in}
\caption{Spectra of the four galaxies in the BRI 1202-0725 system at
  31.25 MHz spectral resolution (= 28 km s$^{-1}$). Figs. 3a-d are:
  the SMG, quasar, Ly$\alpha$-1, and Ly$\alpha$-2, respectively.  The continuum derived
  from the off-line IFs has been subtracted.  The rms noise is 0.9 mJy 
beam$^{-1}$ channel$^{-1}$. Gaussian models have been fit to the spectra,
with parameters as given in Table 1. Note that the low frequency part
of the spectrum of Ly$\alpha$-2 was not sampled by these observations.  
The [CII] line has a rest frequency of 1900.539GHz.
}
\end{figure}

\begin{figure}
\psfig{file=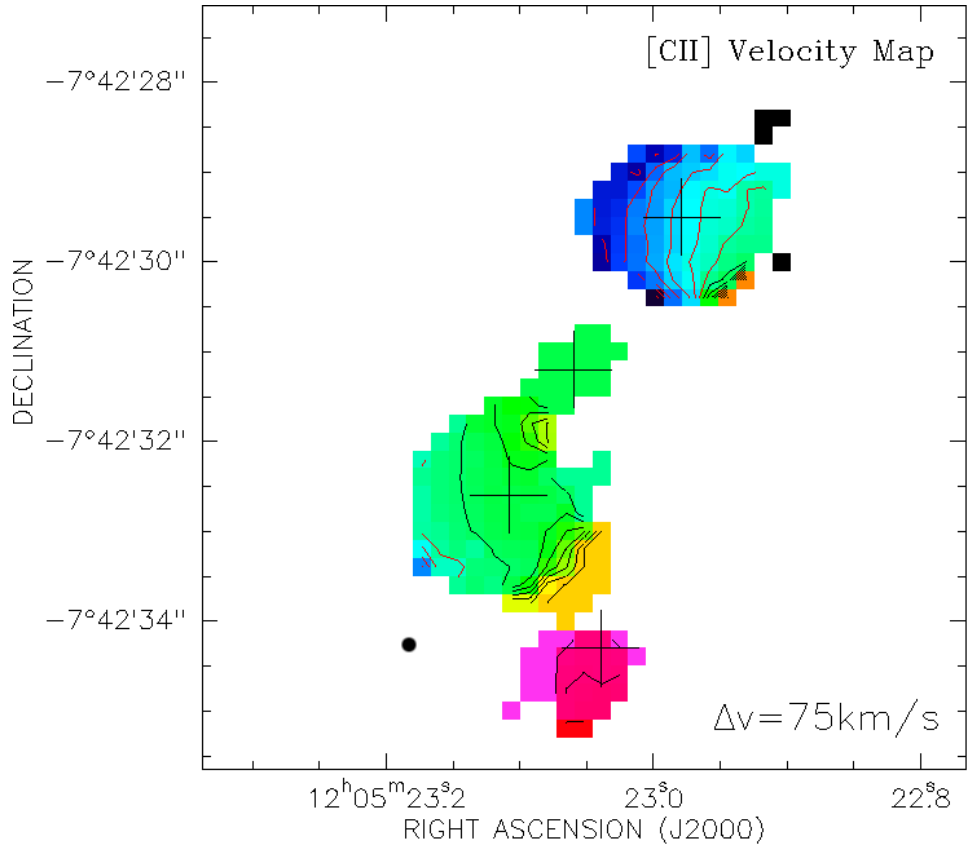,width=5in}
\caption{Isovelocity images (1st moment = intensity weighted mean velocity)
of the [CII] emission from BRI1202-0725, blanked at 3.5$\sigma$.
The color scale ranges from -500 km s$^{-1}$ (blue) to +500 km s$^{-1}$ (red),
and the contours are in steps of 75 km s$^{-1}$. Crosses are the same as Fig. 2.
}
\end{figure}

\begin{figure}
\psfig{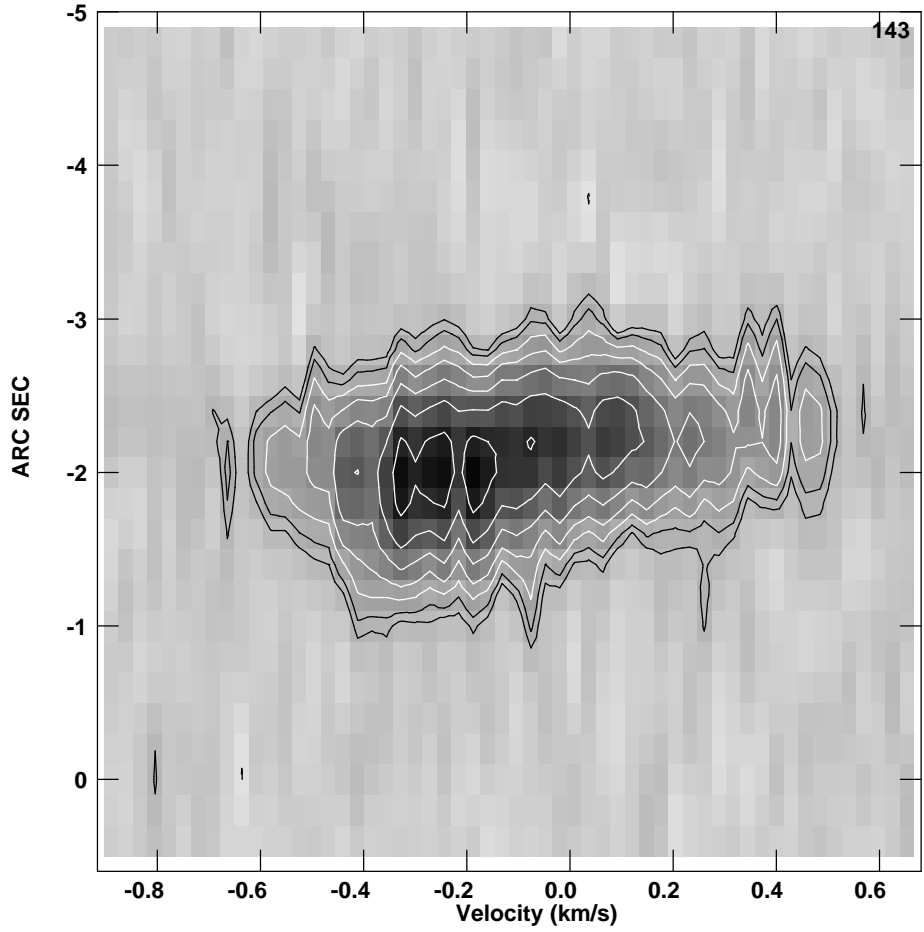}
\caption{ The position-velocity distribution for the SMG in the
BRI1202-0725 system, along the east-west direction. Contour levels are
Contour levels are a geometric progression in square root two,
starting at 2 mJy beam$^{-1}$. Negative contours are dashed.  
Zero velocity corresponds to $z = 4.6941$. }
\end{figure}

\end{document}